\documentclass[draftcls,onecolumn]{IEEEtran}
\usepackage[T1]{fontenc}
\usepackage[latin9]{inputenc}
\usepackage{array}
\usepackage{booktabs}
\usepackage{units}
\usepackage{amsmath}
\usepackage{amssymb}
\usepackage{graphicx}

\makeatletter

\providecommand{\tabularnewline}{\\}

\usepackage{cite}

\usepackage{balance}

\usepackage{amsthm}

\newenvironment{alphafootnotes}
  {\par\edef\savedfootnotenumber{\number\value{footnote}}
   
   \setcounter{footnote}{0}}
  {\par\setcounter{footnote}{\savedfootnotenumber}}

\@ifundefined{showcaptionsetup}{}{%
 \PassOptionsToPackage{caption=false}{subfig}}
\usepackage{subfig}
\makeatother

\begin{document}

\title{High-Dimensional CSI Acquisition in Massive MIMO: Sparsity-Inspired
Approaches}

\author{\IEEEauthorblockN{Juei-Chin Shen, \textit{Member}, \textit{IEEE}, Jun Zhang, \textit{Member}, \textit{IEEE}, Kwang-Cheng Chen, \textit{Fellow}, \textit{IEEE},\\ and Khaled B. Letaief, \textit{Fellow}, \textit{IEEE}}\\}
\maketitle
\begin{abstract}
Massive MIMO has been regarded as one of the key technologies for
5G wireless networks, as it can significantly improve both the spectral
efficiency and energy efficiency. The availability of high-dimensional
channel side information (CSI) is critical for its promised performance
gains, but the overhead of acquiring CSI may potentially deplete the
available radio resources. Fortunately, it has recently been discovered
that harnessing various sparsity structures in massive MIMO channels
can lead to significant overhead reduction, and thus improve the system
performance. This paper presents and discusses the use of sparsity-inspired
CSI acquisition techniques for massive MIMO, as well as the underlying
mathematical theory. Sparsity-inspired approaches for both frequency-division
duplexing and time-division duplexing massive MIMO systems will be
examined and compared from an overall system perspective, including
the design trade-offs between the two duplexing modes, computational
complexity of acquisition algorithms, and applicability of sparsity
structures. Meanwhile, some future prospects for research on high-dimensional
CSI acquisition to meet practical demands will be identified.\begin{alphafootnotes}%
\footnote{J.-C. Shen, J. Zhang, and K. B. Letaief are with the Department of
Electronic and Computer Engineering, Hong Kong University of Science
and Technology, Hong Kong (E-mail: \{eejcshen, eejzhang, eekhaled\}@ust.hk).%
}%
\footnote{K.-C. Chen is with the Graduate Institute of Communication Engineering,
National Taiwan University, Taipei, Taiwan (E-mail: chenkc@cc.ee.ntu.edu.tw).%
}\end{alphafootnotes}\end{abstract}

\section{Introduction}

Massive MIMO systems promise to boost spectral efficiency by more
than one order of magnitude \cite{Marzetta10,Rusek13}. Full benefits
of massive MIMO, however, will never come to fruition without the
base stations (BSs) having adequate channel knowledge, which appears
to be an extremely challenging task \cite{Letaief14}. The challenges
posed by MIMO channels of very high dimension are confronted in both
frequency-division duplexing (FDD) and time-division duplexing (TDD)
massive MIMO systems. In the FDD mode, both the pilot-aided training
overhead and the feedback overhead for CSI acquisition grow proportionally
with the BS antenna size. However, the proportion of radio resources
allocated to CSI acquisition is severely restricted by the channel
coherence period. The situation is made worse in an environment with
high user equipment (UE) mobility.

In view of this, a considerable research effort has been devoted to
TDD massive MIMO by exploiting channel reciprocity. Although the training
overhead for TDD operation becomes proportional to the number of active
UEs rather than that of BS antennas, the inevitable reuse of the same
pilot in neighboring cells can seriously degrade the quality of obtained
channel knowledge. This is because the channels to UEs in adjacent
cells who share the same pilot will be collectively acquired by the
BS. In other words, the desired channel obtained by the BS will be
contaminated by interference channels. Once this contaminated channel
knowledge is utilized for transmitting or receiving data, intercell
interference occurs immediately and hence limits the achievable performance.
This problem, known as \emph{pilot contamination}, can not be circumvented
simply by adding more BS antennas.

\begin{figure*}[t]
\subfloat[]{\noindent \begin{centering}
\includegraphics[width=9cm]{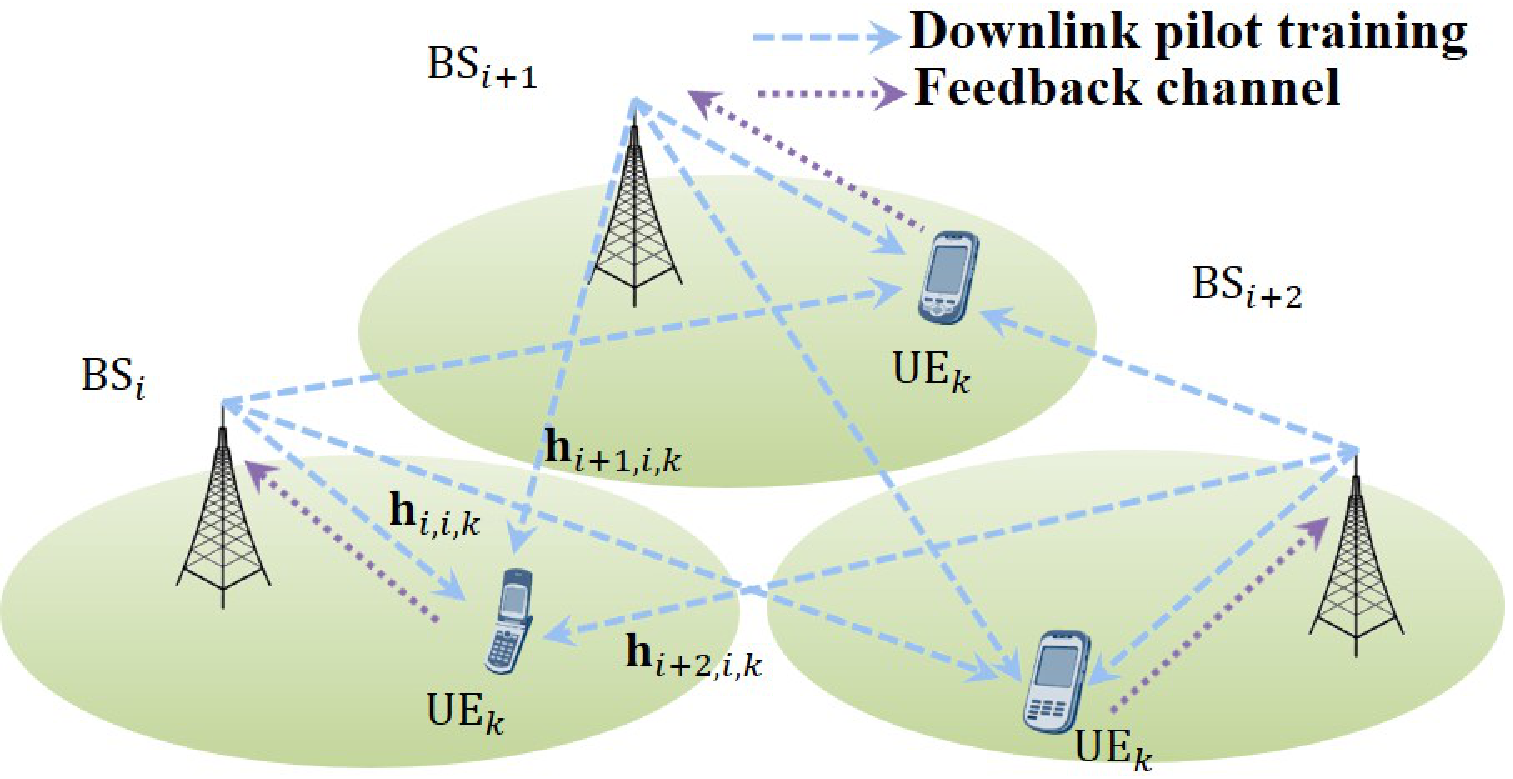}
\par\end{centering}

}\subfloat[]{\noindent \begin{centering}
\includegraphics[width=9cm]{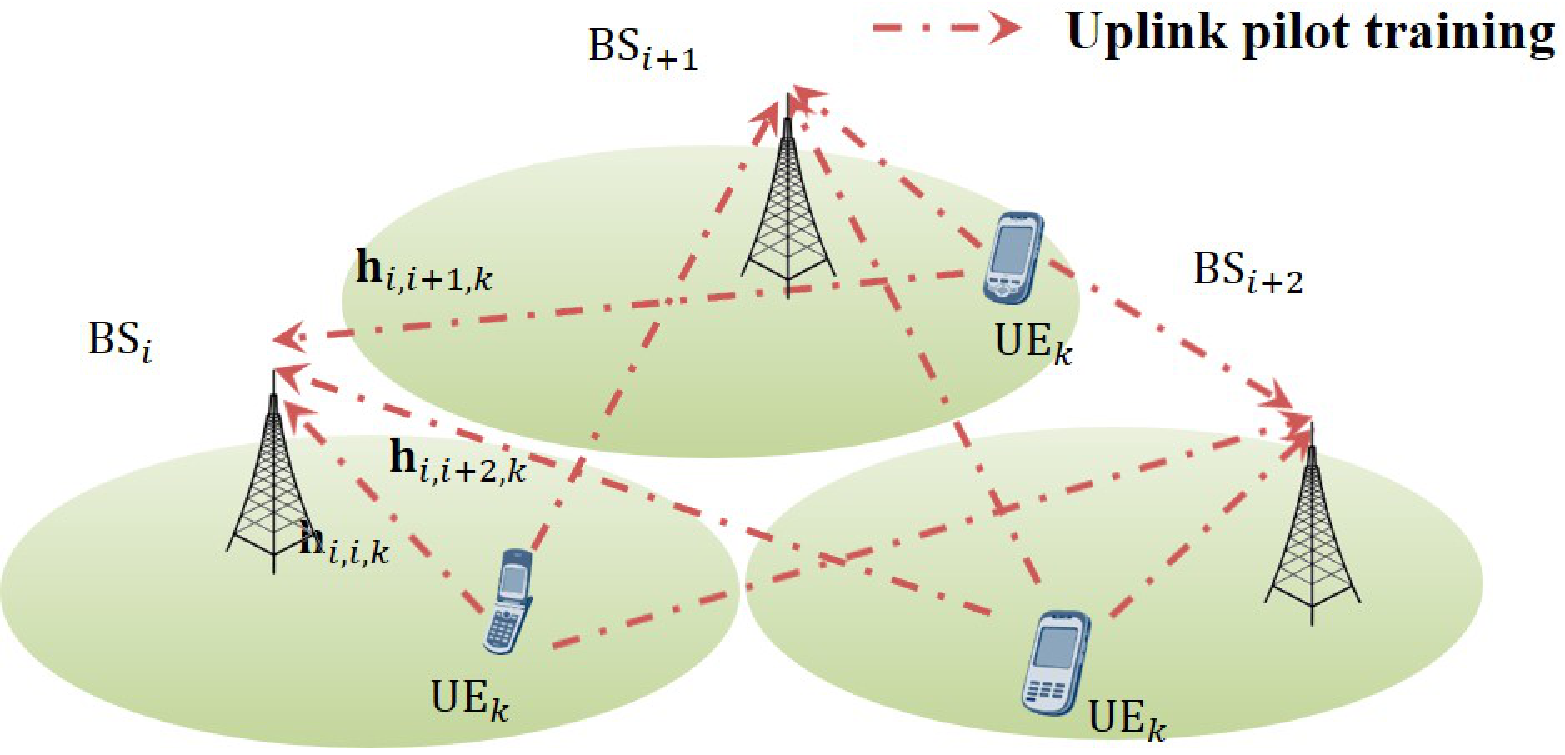}
\par\end{centering}

}

\caption{Pilot reuse in multiple cells. (a) FDD downlink training (b) TDD uplink
training.\label{fig: pilot reuse}}
\end{figure*}

Several attempts have been made to tackle the challenges of acquiring
high-dimensional CSI in massive MIMO. For instance, in \cite{Choi14FDD},
open/closed loop training that utilizes temporal and spatial channel
statistics is proposed to reduce the amount of downlink training overhead.
For mitigating pilot contamination, the optimal design of precoding
matrices aimed at minimizing the square errors caused by pilot reuse
has shown its superiority over linear precoding \cite{Jose11}. Thanks
to the recent advances in compressed sensing \cite{Bajwa10,Eldar12CS},
sparse signal processing has attracted much attention in such high-dimensional
settings, which has also demonstrated its power in CSI acquisition
in terms of reconstructing CSI from a limited number of channel measurements.
Various sparsity structures exhibited by massive MIMO channels have
recently been identified, thereby motivating the development of new
strategies for CSI acquisition. Surprisingly, not only can high training
overhead be reduced, but pilot contamination can also be resolved
by appealing to sparsity-inspired approaches.

In this paper, we provide a comprehensive overview of the state-of-the-art
research on sparsity-inspired approaches for high-dimensional CSI
acquisition. In Section II, the challenges in FDD and TDD massive
MIMO are reviewed in detail, including a rarely mentioned issue of
FDD pilot contamination. On the basis of different sparsity structures,
a variety of methods for either achieving overhead reduction or alleviating
the effects of pilot reuse are examined and compared in Section III.
Finally, concluding remarks are made in Section IV.

\emph{Notations}: $\mathbb{C}$: complex number, $\Re$: real part,
$\left\Vert \cdot\right\Vert _{p}$: $p$-norm, $\left(\cdot\right){}^{'}$:
transpose, $\left(\cdot\right){}^{H}$: Hermitian transpose, $\mathbf{I}_{N}$:
$N\times N$ identity matrix, $\mathcal{N}\left(\cdot,\cdot\right)$:
normal distribution, $\mathbb{E}\left[\cdot\right]$: expectation,
$\mathbf{0}$: zero vector, $\mbox{card}\left(\cdot\right)$: cardinality,
$\mbox{supp}\left(\cdot\right)$: the set of indices of non-zero elements,
$\mbox{Var}\left(\cdot\right)$: variance, $\max\left\{ \cdot\right\} $:
the maximum element, $\mbox{Vec}\left(\cdot\right)$: vectorization,
$\otimes$: Kronecker product, $\succeq$: matrix inequality, $\left(\cdot\right)^{\dagger}$:
pseudo inverse.

\section{Challenges of High-Dimensional CSI Acquisition}

In massive MIMO systems with high-dimensional channels, CSI acquisition
at BSs is a fundamentally challenging problem. In FDD massive MIMO,
performing this task consumes a considerable amount of radio resources
which is proportional to the dimension of channels. On the other hand,
in TDD-mode operation, it is hard to ensure the orthogonality of pilot
sequences in the multicell scenario as the number of overall UEs becomes
large. As a result, the inevitable reuse of correlated pilot sequences
in different cells, known as pilot contamination, causes capacity-limiting
intercell interference. 

To illustrate these difficulties further, we will consider a massive
MIMO network consisting of $L$ hexagonal cells. In each cell, there
is a BS equipped with an $M$-element linear array%
\footnote{For simplicity, the assumption of employing linear arrays is made.
However, most of the results discussed in this paper can be generalized
to include the cases of using planar or cylindrical arrays.%
}, serving $K$ single-antenna UEs. The channel between BS $i$ and
UE $k$ in cell $j$ is denoted by the $M\times1$ vector $\mathbf{h}_{i,j,k}$.
The BS antenna size is supposed to be greatly larger than the number
of served UEs.

\subsection{FDD Massive MIMO}

In the FDD mode, obtaining CSI at BSs is normally performed in two
steps. First, each BS sends a downlink training matrix to its served
UEs. Second, each UE estimates the desired channel based on the downlink
measurements and feeds back acquired CSI through dedicated uplink
feedback channels.

During downlink training, UE $k$ in cell $i$ receives channel measurements

\begin{equation}
\mathbf{y}_{i,k}^{\tiny{\mbox{DL}}}=\mathbf{S}_{i}^{\tiny{\mbox{DL}}}\mathbf{h}_{i,i,k}+\sum_{l\neq i}\mathbf{S}_{l}^{\tiny{\mbox{DL}}}\mathbf{h}_{l,i,k}+\mathbf{z}_{i,k}^{\tiny{\mbox{DL}}},\label{eq: FDD training}
\end{equation}
where $\mathbf{S}_{l}^{\tiny{\mbox{DL}}}$ denotes the $N\times M$
pilot training matrix used in cell $l$, $\mathbf{z}_{i,k}^{\tiny{\mbox{DL}}}$
is the additive noise, while the first term of the right-hand side
(RHS) represents the desired channel measurements, and the next term
results from intercell interference. Even without considering the
impact of intercell interference, the required training overhead $N$
for conventional least-squares (LS) or minimum mean square error (MMSE)
estimators to achieve a reasonable performance level still scales
linearly with the BS antenna size. By taking intercell interference
into account, a further increase in training overhead would occur.
The explicit expressions of the optimal pilot training matrices ($N\geq M$)
are provided in \cite{Biguesh06} for single-cell networks. In \cite{Kang11pilot},
the optimal design of training matrices for multicell MIMO-OFDM systems
is considered. 

What makes the situation worse is that typical feedback channels are
finite-rate. This implies that only quantized versions of channel
estimates can be fed back to BSs. If there are predefined codebooks
consisting of precoding vectors, then the index of the optimal codebook
vector is required to be sent back \cite{Love08feedback,Ghosh10LTE}.
However, either the amount of quantized CSI or the size of codebooks
increases in proportion to the number of BS antennas, and it in turn
makes these two limited feedback techniques impractical in FDD massive
MIMO.

Note that when the same training matrix is repeatedly used in multiple
cells, i.e., $\mathbf{S}_{1}^{\tiny{\mbox{DL}}}=\cdots=\mathbf{S}_{L}^{\tiny{\mbox{DL}}}$,
this can be regarded as pilot contamination in FDD massive MIMO. As
a result of such contamination, as shown in Fig. \ref{fig: pilot reuse}(a),
BS $i$ will acquire the composite channel $\sum_{l=1}^{L}\mathbf{h}_{l,i,k}$
rather than the desired channel $\mathbf{h}_{i,i,k}$, given the feedback
channel being error-free and the additive noise being ignored. Despite
this fact, utilizing this composite CSI to form a precoding vector
and transmit signals at BS $i$ will not cause serious interference
to UEs in the neighboring cells. For instance, given that maximum
ratio transmission (MRT) precoding is employed, the transmitted signal
from BS $i$ can be expressed as $\mathbf{x}_{i}=\sum_{k=1}^{K}\mathbf{w}_{i,k}^{\tiny{\mbox{FDD}}}x_{i,k}$
where $x_{i,k}$ is the signal intended for UE $k$ within the cell,
and $\mathbf{w}_{i,k}^{\tiny{\mbox{FDD}}}=\sum_{l=1}^{L}(\mathbf{h}_{l,i,k}^{H})^{'}$
denotes the MRT precoding vector. During the downlink transmission
phase, the received interference at UE $m$ in cell $j$ due to BS
$i$ is given by

\begin{align}
I_{i,j,m} & =\mathbf{h}_{i,j,m}^{'}\mathbf{x}_{i}=\sum_{k=1}^{K}\sum_{l=1}^{L}\mathbf{h}_{l,i,k}^{H}\mathbf{h}_{i,j,m}x_{i,k}.
\end{align}
When the number of BS antennas grows without limit, the channel vectors
are asymptotically orthogonal. Thus, the channel products $\mathbf{h}_{l,i,k}^{H}\mathbf{h}_{i,j,m}$
approach zero and so does the interference $I_{i,j,m}$. In other
words, intercell interference caused by pilot contamination diminishes
asymptotically with increasing BS antenna size. This implies that
there is no need to mitigate intercell interference by making training
matrices distinct from each other in the asymptotic regime. Hence,
the existing literature rarely addresses the issue of pilot contamination
in FDD massive MIMO. 

Note that uplink training in the FDD mode is not considered here.
An explanation for this is provided as follows. The uplink CSI is
mainly utilized for data acquisition in a multiple-access channel,
instead of a broadcast channel. This means that more advanced signal
processing techniques, such as blind multiuser detection, can be applied
at the BS side. Thus, pilot-aided training may not be the best choice
and CSI acquisition is not necessarily separated from data acquisition.

\subsection{TDD Massive MIMO \label{sub: Challenges in TDD massive MIMO}}

Making massive MIMO operate in the TDD mode is a promising way to
circumvent the identified difficulties in the FDD mode. Owing to channel
reciprocity in the TDD mode, the CSI obtained via uplink training
can be utilized for downlink transmission. More importantly, the cost
of uplink training now increases linearly with the number of active
UEs rather than that of BS antennas. Typically, for obtaining accurate
CSI, it requires that each UE transmits an orthogonal pilot sequence
to its serving BS. However, the number of available orthogonal pilot
sequences is limited by the ratio of the channel coherence interval
to the channel delay spread \cite{Larsson14}, which may be small
due to the mobility of UEs or adverse physical environments. When
the number of overall UEs becomes large, the situation of using non-orthogonal
pilot sequences, known as pilot contamination, inevitably arises.
A consequence of pilot contamination is intra- and inter-cell interference.

\begin{figure}
\begin{centering}
\includegraphics[width=9cm,height=6cm]{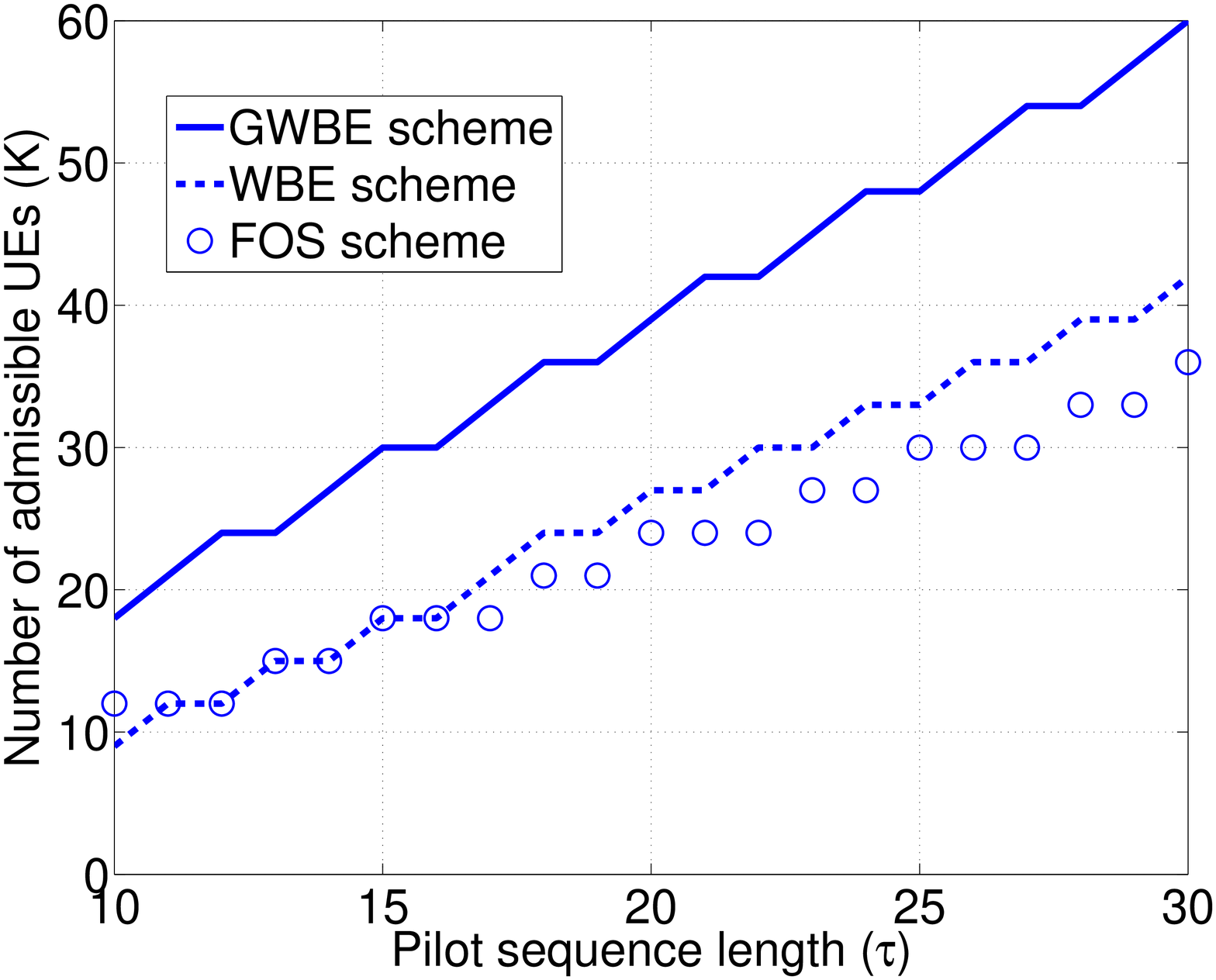}
\par\end{centering}

\caption{The number ($K=3l$) of admissible UEs versus pilot sequence length
for the GWBE, WBE, and FOS schemes, given a fixed SINR-requirement
pattern, that is $\left\{ \gamma_{1\sim l}=\nicefrac{1}{3},\mbox{ }\gamma_{\left(l+1\right)\sim2l}=1,\mbox{ }\gamma_{\left(2l+1\right)\sim3l}=3\right\} $
(from \cite{Shen14Globecom}.) \label{fig: UEs versus Tau}}
\end{figure}

During the uplink training phase, the received signal at the $i$th
BS is given by

\begin{equation}
\mathbf{Y}_{i}^{\tiny{\mbox{UL}}}=\sum_{l=1}^{L}\mathbf{S}_{l}^{\tiny{\mbox{UL}}}\mathbf{H}_{i,l}+\mathbf{Z}_{i}^{\tiny{\mbox{UL}}},\label{eq: uplink training}
\end{equation}
where $\mathbf{H}_{i,l}=[\mathbf{h}_{i,l,1},\ldots,\mathbf{h}_{i,l,K}]^{'}$
consists of channel vectors from UEs in the $l$th cell to the $i$th
BS, the columns of $\mathbf{S}_{l}^{\tiny{\mbox{UL}}}$ form a set
of $\tau\times1$ pilot sequences $\{\mathbf{s}_{l,k}\}_{k=1}^{K}$,
and $\mathbf{Z}_{i}^{\tiny{\mbox{UL}}}$ denotes an additive noise
matrix. To illustrate the case of intercell interference, assume that
the same set of orthogonal pilot sequences is reused in each cell,
i.e., $\mathbf{S}_{1}^{\tiny{\mbox{UL}}}=\cdots=\mathbf{S}_{L}^{\tiny{\mbox{UL}}}$
and $\mathbf{s}_{l,k_{1}}^{'}\mathbf{s}_{l,k_{2}}=0$ for $k_{1}\neq k_{2}$,
as shown in Fig. \ref{fig: pilot reuse}(b). Employing the LS estimator
yields the channel estimate

\begin{eqnarray}
\mathbf{\widehat{H}}_{i,i} & = & \left[\left(\mathbf{S}_{i}^{\tiny{\mbox{UL}}}\right)^{H}\mathbf{S}_{i}^{\tiny{\mbox{UL}}}\right]\left(\mathbf{S}_{i}^{\tiny{\mbox{UL}}}\right)^{H}\mathbf{Y}_{i}^{\tiny{\mbox{UL}}},\nonumber \\
 & = & \mathbf{H}_{i,i}+\sum_{l\neq i}\mathbf{H}_{i,l}+\left[\left(\mathbf{S}_{i}^{\tiny{\mbox{UL}}}\right)^{H}\mathbf{S}_{i}^{\tiny{\mbox{UL}}}\right]\left(\mathbf{S}_{i}^{\tiny{\mbox{UL}}}\right)^{H}\mathbf{Z}_{i}^{\tiny{\mbox{UL}}},\label{eq: LS estimate}
\end{eqnarray}
where the rows of $\mathbf{\widehat{H}}_{i,i}$ are given by $\mathbf{\hat{h}}_{i,i,k}=\sum_{l=1}^{L}\mathbf{h}_{i,l,k}$
when ignoring the noise. During downlink transmission, using estimates
$\mathbf{\hat{h}}_{i,i,k}$ to form the transmit signal $\mathbf{x}_{i}=\sum_{k=1}^{K}\mathbf{w}_{i,k}^{\tiny{\mbox{TDD}}}x_{i,k}$,
where $\mathbf{w}_{i,k}^{\tiny{\mbox{TDD}}}=\sum_{l=1}^{L}(\mathbf{h}_{i,l,k}^{H})^{'}$
are MRT precoding vectors, will cause interference

\begin{eqnarray}
I_{i,j,m} & = & \mathbf{h}_{i,j,m}^{'}\mathbf{x}_{i}\nonumber \\
 & = & \left\Vert \mathbf{h}_{i,j,m}\right\Vert _{2}^{2}x_{i,m}+\sum_{\substack{k\neq m\\
\tiny{\mbox{or }}l\neq j
}
}\mathbf{h}_{i,l,k}^{H}\mathbf{h}_{i,j,m}x_{i,k}\label{eq: TDD inteference}
\end{eqnarray}
to UE $m$ in cell $j$. Though the second term on the RHS of (\ref{eq: TDD inteference})
decreases with the increasing BS antenna size, the first term, which
does not vanish, makes the received signal-to-interference-plus-noise
ratio (SINR) at UE $m$ in cell $j$ converge to a limit and becomes
the performance limiting factor.

The current investigation into TDD pilot contamination focuses on
its impact on the received SINR or the sum rate when linear precoders/detectors
are applied. However, very little is known about its impact on the
system equipped with nonlinear precoders/detectors. A recent work
\cite{Shen14Globecom} provides an interesting perspective on the
user capacity of pilot-contaminated massive MIMO which quantifies
the maximum number of admissible UEs given their own SINR requirements.
As shown in Fig. \ref{fig: UEs versus Tau}, the user capacity of
three schemes%
\footnote{The pilot sequences employed in the GWBE, WBE, and FOS schemes are
respectively generalized Welch bound equality (GWBE) sequences, WBE
sequences, and finite orthogonal sequences (FOS) whose correlation
among sequences is either 1 or 0. The same downlink power allocation,
$P_{i}\propto\nicefrac{\gamma_{i}}{(1+\gamma_{i})}$, is used in the
three schemes.%
} of joint pilot design and transmit power allocation is fundamentally
limited by the length of pilot sequences. For further details about
pilot contamination in TDD massive MIMO, the study \cite{Lu14overview}
and references therein should be consulted.

\section{Sparsity-Inspired CSI Acquisition}

Despite the challenges imposed by the high dimensionality of channel
matrices, a number of research efforts have sought to address them
and have achieved reasonably efficient CSI acquisition. In particular,
sparsity-inspired approaches have been proved to be powerful tools,
as presented below.

\subsection{FDD Massive MIMO}

\subsubsection{The Joint CSI Recovery Method}

Authors of \cite{Rao14} proposed a method for low-overhead pilot
training in the single-cell scenario, taking advantage of channel
sparsity. Provided that a uniform linear array with critically spaced
antennas is employed at the BS, the channel $\mathbf{h}_{k}$, where
indices of BSs are discarded in the single-cell scenario, exhibits
a sparse representation $\mathbf{h}_{k}^{\mathrm{a}}$ in the angular
domain, i.e., 

\begin{equation}
\mathbf{h}_{k}=\mathbf{U}\mathbf{h}_{k}^{\mathrm{a}},
\end{equation}
where $\mathbf{U}$ is a discrete Fourier transform (DFT) matrix whose
columns form an angular basis. The cardinality of $\mbox{supp}(\mathbf{h}_{k}^{\mathrm{a}})$
can be reasonably assumed to be greatly less than $M$ because of
limited local scattering at the BS whose antenna array mounted higher
than surrounding scatterers. Additionally, based on the results in
\cite{Poutanen10}, it has been argued that the channels to UEs are
likely to share a partially common support in the angular domain,
i.e., $\cap_{k=1}^{K}\mbox{supp}(\mathbf{h}_{k}^{\mathrm{a}})=\Omega_{c}$.
In order to utilize the channel sparsity and common support property
simultaneously, channel measurements acquired at UEs are fed back
to the serving BS via error-free feedback channels. Hence, a joint
channel recovery problem can be formulated as follows:

\begin{equation}
\begin{array}{cc}
\underset{\left\{ \mathbf{h}_{k},\forall k\right\} }{\min} & \sum_{k=1}^{K}\left\Vert \mathbf{y}_{k}^{\tiny{\mbox{DL}}}-\mathbf{S}^{\tiny{\mbox{DL}}}\mathbf{h}_{k}\right\Vert _{1}^{2}\\
\mbox{s.t.} & \cap_{k=1}^{K}\mbox{supp}(\mathbf{h}_{k}^{\mathrm{a}})=\Omega_{c}.
\end{array}
\end{equation}
Using orthogonal matching pursuit (OMP) as a basis, a greedy algorithm
has been proposed to efficiently solve this problem. The simulation
results show that the required training overhead for this recovery
algorithm can be significantly less than that for the conventional
LS estimator. Moreover, the mean square error (MSE) performance improves
with the increasing cardinality of $\Omega_{c}$. 

One major concern about this joint recovery approach is the underlying
assumption of perfect channel measurements being fed back. As practical
feedback channels are rate-limited, it is more reasonable to assume
quantized measurements at the BS. The impact of quantization on the
channel recovery performance requires further investigation. On the
other hand, it has been suggested that the amount of channel measurements
that is needed at the BS should be adaptively adjusted according to
the sensitivity of the system performance to the CSI inaccuracy \cite{Kuo12}.
Furthermore, there has been little quantitative analysis of the required
training overhead against the channel sparsity level. This quantification
is in dire need as it will help us measure the actual training overhead
reduction that can be achieved without relying on time-consuming simulations.

\subsubsection{The Weighted $\ell_{1}$ Minimization Method}

\begin{figure}
\noindent \begin{centering}
\includegraphics[width=9cm,height=6cm]{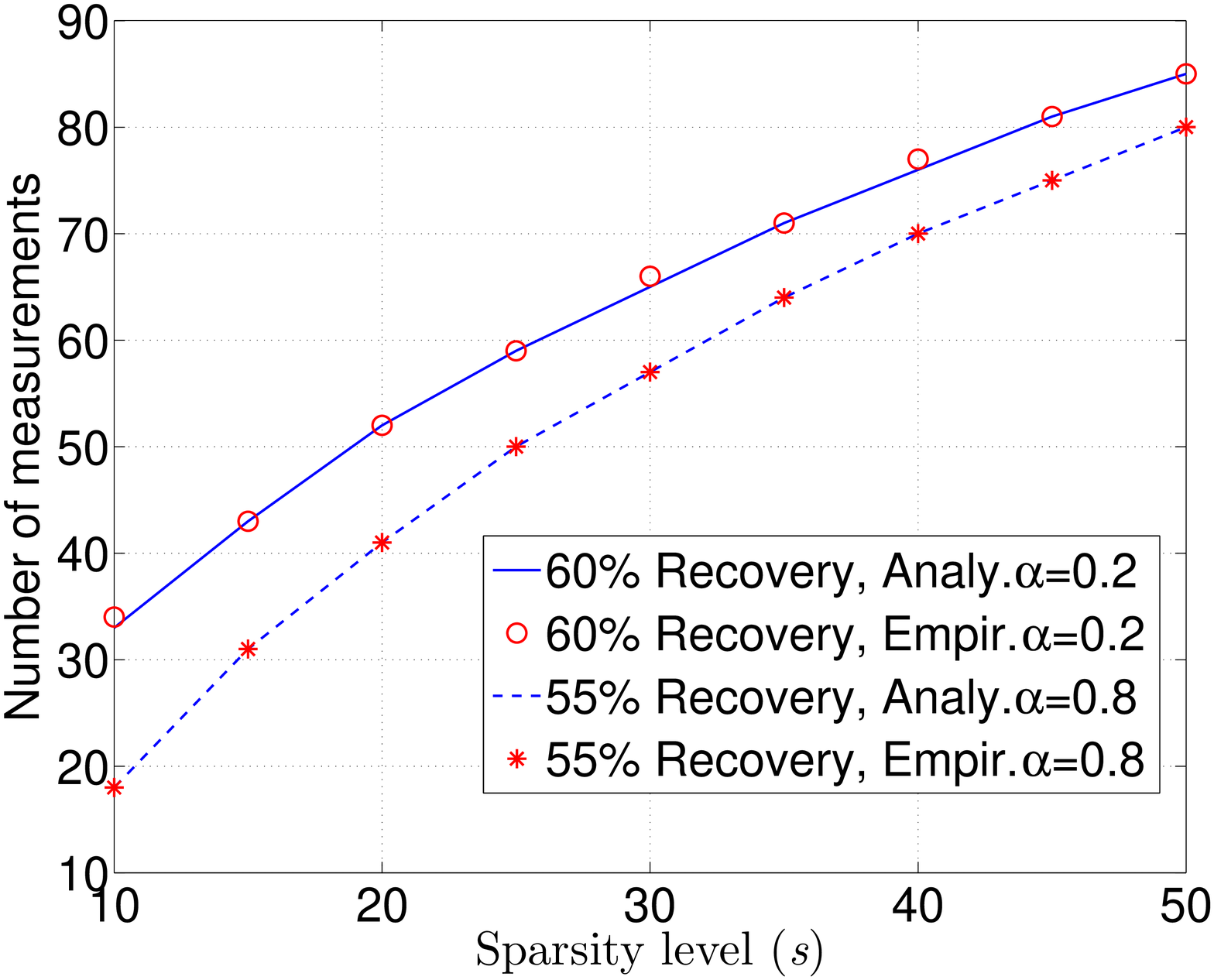}
\par\end{centering}

\caption{Phase transition curves of (\ref{eq: weighted l1 min}) over different
values of $\alpha$ given $M=100$, $\hat{s}=10$, $\mathbf{z}_{k}^{\tiny{\mbox{DL}}}=\mathbf{0}$,
and $\epsilon=0$ (from \cite{Shen14CS}.)\label{fig: Phase-transition without error 02}}
\end{figure}

Considering a similar single-cell scenario, the study in \cite{Shen14CS}
has drawn attention to utilizing partial support information of sparse
massive MIMO channels, which is a collection of indices of significant
entries of channel vectors in the angular domain. The main advantage
of using partial support information is the possibility of achieving
a remarkable training overhead reduction. Specifically, the order
of the required overhead decreases from $\mathcal{O}\left(s\log M\right)$
to $\mathcal{O}\left(s\right)$ where $s=\mbox{card}[\mbox{supp}(\mathbf{h}_{k}^{\mathrm{a}})]$
is the channel sparsity level. Assume that the partial support information
$\widehat{T}_{k}$ of channel $\mathbf{h}_{k}^{\mathrm{a}}$ is available
at UE $k$, where $\mbox{card}(\widehat{T}_{k})=\hat{s}$ and $\mbox{card}[\mbox{supp}(\mathbf{h}_{k}^{\mathrm{a}})\cap\widehat{T}_{k}]$
is given by $\left\lfloor \alpha\hat{s}\right\rfloor $. The higher
the factor $\alpha$, the higher is the accuracy level of partial
support information. Based on a weighted $\ell_{1}$ minimization
framework, the channel recovery is performed as follows: 

\begin{equation}
\begin{array}{cc}
\underset{\mathbf{\hat{h}}_{k}^{\mathrm{a}}\in\mathbb{\mathbb{C}}^{M}}{\min} & \left\Vert \mathbf{\hat{h}}_{k}^{\mathrm{a}}\right\Vert _{1,\mathbf{w}}\\
\mbox{subject to} & \left\Vert \mathbf{S}^{\tiny{\mbox{DL}}}\mathbf{U}\mathbf{\hat{h}}_{k}^{\mathrm{a}}-\mathbf{y}_{k}^{\tiny{\mbox{DL}}}\right\Vert _{2}\leq\epsilon,\\
\mbox{with} & w_{i}=\begin{cases}
1, & i\notin\widehat{T}_{k},\\
0, & i\in\widehat{T}_{k},
\end{cases}\mbox{ }
\end{array}\label{eq: weighted l1 min}
\end{equation}
where $\mathbf{S}^{\tiny{\mbox{DL}}}\in\mathbb{C}^{N\times M}$ is
designed to be a Gaussian random matrix of independent complex normal
entries, the noise $\mathbf{z}_{k}^{\tiny{\mbox{DL}}}$ is assumed
to be upper bounded, i.e., $\left\Vert \mathbf{z}_{k}^{\tiny{\mbox{DL}}}\right\Vert _{2}\leq\epsilon$,
and $||\mathbf{\hat{h}}_{k}^{\mathrm{a}}||_{1,\mathbf{w}}=\sum_{i=1}^{M}w_{i}|\hat{h}_{k}^{\mathrm{a}}[i]|$.
In the objective function, the entries that are expected to be zero
are weighted more heavily than others. The results show a significant
improvement over the method without using partial support information
when the accuracy level $\alpha$ exceeds a certain threshold. Moreover,
taking a convex geometry approach, the authors have successfully and
precisely quantified the required training overhead for achieving
a certain percentage of exact recovery. The exact recovery is declared
if $||\mathbf{\hat{h}}_{k}^{\mathrm{a}}-\mathbf{h}_{k}^{\mathrm{a}}||_{2}\leq10^{-4}$.
As shown in Fig. \ref{fig: Phase-transition without error 02}, the
analytical curves of $\alpha=0.2$ and $\alpha=0.8$ can accurately
depict the empirical phase transition curves of $60\%$ exact recovery
and $55\%$ exact recovery, respectively.

Unlike the previous method, here, channel measurements are not fed
back to the BS. In other words, it avoids the assumption of error-free
feedback channels. However, it raises another issue of storing random
matrices at UEs with limited memory. Also, performing convex optimization
can impose a stringent computation requirement on UEs without seeking
for low-complexity solutions. Several attempts have been made to design
practical training matrices. In \cite{Bajwa07Toeplitz}, Toeplitz-structured
training matrices, suggested for the realistic implementation, are
shown to perform comparably to Gaussian random matrices and require
generating less independent random variables. A deterministic approach
to the training matrix design is first considered by appealing to
matrix properties such as mutual coherence \cite{Elad07Porj}. More
advanced deterministic training matrices are developed in \cite{Li13CSmatrix}
to yield higher recovery accuracy. In the context of FDD massive MIMO,
it would be interesting to invent structurally random or deterministic
training matrices that take partial support information of channels
to multiple UEs into consideration. In addition, the similar concepts
of using prior channel knowledge to lower training overhead can be
found in \cite{Choi14FDD} where spatial and temporal correlations
are harnessed. More study is needed to better understand how to integrate
all the relevant prior knowledge into efficient CSI acquisition.

\subsection{TDD Massive MIMO}

As mentioned in Sec. \ref{sub: Challenges in TDD massive MIMO}, employing
uplink training to obtain high-dimensional downlink CSI results in
undesired pilot contamination, and the following are some efforts
to address this issue.

\subsubsection{The Coordinated MMSE Method}

Contradicting conventional wisdom, it has been shown that it is possible
to mitigate pilot contamination using the linear MMSE estimator \cite{Yin13}.
The key factor in determining the success of MMSE estimation is that
each channel to the UE can be regarded as a linear combination of
finite steering vectors

\begin{equation}
\mathbf{h}_{i,j,k}=\frac{1}{\sqrt{P}}\sum_{p=1}^{P}\alpha_{i,j,k}\left(p\right)\mathbf{a}\left[\theta_{i,j,k}\left(p\right)\right],\label{eq: multipath model}
\end{equation}
where $P$ is the number of paths, $\alpha_{i,j,k}\left(p\right)$
are zero-mean path gains, and $\mathbf{a}\left[\theta_{i,j,k}\left(p\right)\right]$
denote the steering vectors due to angle of arrivals (AoAs) $\theta_{i,j,k}\left(p\right)$.
Consequently, the rank of the channel covariance matrix $\mathbf{R}_{i,j,k}\triangleq\mathbb{E}\{\mathbf{h}_{i,j,k}\mathbf{h}_{i,j,k}^{H}\}$
depends on the range $[\theta_{i,j,k}^{\min},\theta_{i,j,k}^{\max}]$
in which AoAs $\theta_{i,j,k}\left(p\right)$ lie, which typically
turns out to be low. Let us focus on the $k$th row of (\ref{eq: LS estimate}),
i.e., $\mathbf{\hat{h}}_{i,i,k}=\sum_{l=1}^{L}\mathbf{h}_{i,l,k}+\mathbf{z}_{i,k}$.
Based on it, the desired channel $\mathbf{h}_{i,i,k}$ can be further
extracted by the MMSE estimator, i.e.,

\begin{eqnarray}
\mathbf{\hat{\hat{h}}}_{i,i,k} & = & \mathbf{R}_{i,i,k}\left(\sigma_{z}^{2}\mathbf{I}_{M}+\sum_{l=1}^{L}\mathbf{R}_{i,l,k}\right)^{-1}\mathbf{\hat{h}}_{i,i,k},
\end{eqnarray}
where the covariance matrix of $\mathbf{z}_{i,k}$ is assumed to be
$\sigma_{z}^{2}\mathbf{I}_{M}$. When the range of AoAs due to interfering
UEs that use the same pilot sequence does not overlap with the AoA
range due to the desired UE, the estimate $\mathbf{\hat{\hat{h}}}_{i,i,k}$
approaches the desired $\mathbf{h}_{i,i,k}$ as the BS antenna size
grows to infinity. This feature is highly attractive because the dimension
of the BS antennas can be made as large as desired in massive MIMO.
Moreover, the condition of non-overlapping AoA ranges can be satisfied
if the reused pilot sequence is properly allocated to UEs in neighboring
cells. A heuristic algorithm has been developed to perform pilot allocation
in a coordinated manner. Another favorable feature of this method
recently demonstrated in \cite{Yin14} is that the asymptotically
optimal estimate is obtainable whether uniform or non-uniform arrays
are employed. As a result, BS antenna arrays are exempt from the requirement
of high calibration accuracy.

The second-order statistics of high-dimensional channels have successfully
been utilized to facilitate robust MMSE channel estimation under pilot
contamination. However, obtaining channel covariance matrices of high
dimension imposes another challenge to the massive MIMO system. It
is interesting to know if the low-rankness can help speed up the acquisition
of channel covariance matrices. Furthermore, it is still unknown if
this covariance-matrix-aware method is sensitive to the inaccuracy
of the second-order statistics. On the other hand, the information
about AoAs actually can be extracted from statistical channel knowledge
prior to commencing the instantaneous CSI acquisition \cite{Foutz2008DOA}.
In this case, the dimension of the parameter space of each channel
shrinks to $P$, which can be significantly less than the original.
Most importantly, this information could aid BSs in distinguishing
between training signals from UEs using the same pilot. 

\begin{table*}[t]
\caption{Comparison of Sparsity-Inspired CSI Acquisition Methods\label{tab: Comparison}}

\centering{}%
\begin{tabular}{>{\raggedright}m{2.5cm}>{\raggedright}m{3cm}>{\raggedright}m{4.5cm}>{\raggedright}m{4.5cm}}
\toprule 
Methods & Sparsity Types & Pros & Cons\tabularnewline
\midrule
\midrule 
Joint CSI Recovery (FDD) & Sparse channel vectors \& Common supports & \begin{itemize}
\item Jointly exploit sparsity \& common-support property
\item Perform channel recovery at the BS\end{itemize}
 & \begin{itemize}
\item UEs need to feed back perfect channel measurements\end{itemize}
\tabularnewline
\midrule 
Weighted $\ell_{1}$Minimization (FDD) & Sparse channel vectors \& partial support information & \begin{itemize}
\item Sharp estimate of the required training overhead
\item Lower training overhead\end{itemize}
 & \begin{itemize}
\item Need to obtain partial support information\end{itemize}
\tabularnewline
\midrule 
Coordinated MMSE (TDD) & Low-rank channel covariance matrices & \begin{itemize}
\item Performance improves with increasing antenna size
\item Lower training overhead\end{itemize}
 & \begin{itemize}
\item Need to obtain second-order channel statistics\end{itemize}
\tabularnewline
\midrule 
Quadratic SDP (TDD) & Low-rank channel matrices & \begin{itemize}
\item No need for knowledge of second-order channel statistics\end{itemize}
 & \begin{itemize}
\item Only suitable for poor scattering propagation environments
\item Higher training overhead\end{itemize}
\tabularnewline
\midrule 
Sparse Bayesian Learning (TDD) & Sparse channel vectors in the UE domain & \begin{itemize}
\item No need for knowledge of second-order channel statistics\end{itemize}
 & \begin{itemize}
\item Channels are not jointly recovered
\item Higher training overhead\end{itemize}
\tabularnewline
\bottomrule
\end{tabular}
\end{table*}

\subsubsection{The Quadratic Semidefinite Programming (SDP) Method}

It is suggested that a BS should collect CSI of both the desired links
within the cell and interference links from its neighboring cells
\cite{Nguyen13LowRank}. In other words, the CSI of interference links
should not be regarded as irrelevant information. From this new angle,
the expression (\ref{eq: uplink training}) can be recast as 

\begin{equation}
\mathbf{Y}_{i}^{\tiny{\mbox{UL}}}=\mathbf{\underline{S}}^{\tiny{\mbox{UL}}}\mathbf{H}_{i}+\mathbf{Z}_{i}^{\tiny{\mbox{UL}}},
\end{equation}
where $\mathbf{\underline{S}}^{\tiny{\mbox{UL}}}\triangleq[\mathbf{S}_{1}^{\tiny{\mbox{UL}}},\ldots,\mathbf{S}_{L}^{\tiny{\mbox{UL}}}]$
and $\mathbf{H}_{i}\triangleq[\mathbf{H}_{i,1}^{'},\ldots,\mathbf{H}_{i,L}^{'}]^{'}$
is the full CSI of wireless links that should be recovered. Thus,
the currently challenging issue is similar to that in FDD massive
MIMO, i.e., how to reduce the required training overhead. 

In the undesirable scattering propagation environments, the rank of
the channel matrix is equal to the number $r$ of the feasible AoAs
$\theta_{i,j,k}\left(p\right)$ in (\ref{eq: multipath model}), which
is greatly less than $\max\left\{ M,K\cdot L\right\} $. Based on
this observation, a unclear norm regularized problem can be formulated
as 

\begin{equation}
\begin{array}{cc}
\underset{\mathbf{H}_{i}}{\min} & \frac{1}{2}\left\Vert \mbox{vec}\left(\mathbf{Y}_{i}^{\tiny{\mbox{UL}}}\right)-\Psi\mbox{vec}\left(\mathbf{H}_{i}\right)\right\Vert _{2}^{2}+\gamma\left\Vert \mathbf{H}_{i}\right\Vert _{F}\end{array},
\end{equation}
where $\Psi=\mathbf{\underline{S}}^{\tiny{\mbox{UL}}}\otimes\mathbf{I}_{M}$
and $\gamma$ is a regularization factor. The sole purpose of adopting
unclear norm regulation is to minimize the sum of the matrix's singular
values, thereby achieving rank minimization. The above problem has
been further recast as a quadratic SDP problem

\begin{equation}
\begin{array}{cc}
\underset{\mathbf{v}}{\min} & \frac{1}{2}\mathbf{v}^{H}\mathbf{v}-\Re\left\{ \left[\mbox{vec}\left(\mathbf{Y}_{i}^{\tiny{\mbox{UL}}}\right)\right]^{H}\mathbf{v}\right\} \\
\mbox{s.t.} & \left[\begin{array}{cc}
\gamma\mathbf{I}_{KL} & \mbox{vec}_{KL,M}^{-1}\left(\Psi^{H}\mathbf{v}\right)\\
\left[\mbox{vec}_{KL,M}^{-1}\left(\Psi^{H}\mathbf{v}\right)\right]^{H} & \gamma\mathbf{I}_{M}
\end{array}\right]\succeq0.
\end{array}
\end{equation}
The solution $\mathbf{v}^{*}$ to this SDP problem determines the
estimate of the channel matrix

\begin{equation}
\mathbf{H}_{i}^{*}=\mbox{vec}_{KL,M}^{-1}\left\{ \Psi^{\dagger}\left[\mbox{vec}\left(\mathbf{Y}_{i}^{\tiny{\mbox{UL}}}\right)-\mathbf{v}^{*}\right]\right\} ,
\end{equation}
which can now be obtained efficiently, thanks to the readily available
polynomial-time SDP solvers.

In the commencing study of massive MIMO \cite{Marzetta06}, the CSI
of interference links at BSs is viewed as nonessential. This is because
that desired links and interference links are asymptotically orthogonal,
and more importantly, intercell interference can be proved manageable
with the CSI of desired links only. Here, we offer an explanation
why there is a need for acquiring the CSI of interference links in
the poor scattering environments. Consider that $\mathbf{H}_{i}=\mathbf{G}_{i}\mathbf{A}$
where $\mathbf{A}=[\mathbf{a}\left(\phi_{1}\right),\ldots,\mathbf{a}\left(\phi_{r}\right)]^{'}$
is an $r\times M$ matrix of full row rank with $r\ll\min\left\{ M,KL\right\} $
due to poor scattering, and $\mathbf{G}_{i}$ consists of $KL\times r$
independent and identically distributed (i.i.d.) zero-mean channel
gains. Then, we have $\lim_{M\rightarrow\infty}\mathbf{A}\mathbf{A}^{H}=\mathbf{I}_{r}$
and

\begin{equation}
\lim_{M\rightarrow\infty}\mathbf{H}_{i}\mathbf{H}_{i}^{H}=\mathbf{G}_{i}\mathbf{G}_{i}^{H}\not\propto\mathbf{I}_{KL}
\end{equation}
which implies that the correlation among wireless links does not diminish
with the increasing BS antenna size. In such a situation, it becomes
crucial to obtain the full CSI of wireless links for effective interference
management.

\subsubsection{The Sparse Bayesian Learning (SBL) Method}

Sharing the same perspective as the study \cite{Nguyen13LowRank},
the work in \cite{Wen14} also considers acquiring the full CSI of
wireless links and proposes a sparse Bayesian learning method to achieve
this goal. Sparse Bayesian learning was first presented in \cite{Tipping02Bayesian}
and has been proved to outperform some prevailing $\ell_{1}$ minimization
algorithms \cite{Zhang11Bayesian}. The SBL method proceeds by first
transforming the channel matrix into the angular domain via DFT as
mentioned in the joint CSI recovery method, i.e., $\mathbf{\underline{H}}_{i}=\mathbf{H}_{i}\mathbf{U}$.
Interestingly, instead of taking advantage of the sparsity in the
angular domain, the sparsity in the UE domain, which has been empirically
shown to exist, is utilized. In other words, the column vectors of
the channel matrix $\mathbf{\underline{H}}_{i}$ are considered one
by one. As each column vector consists of elements due to different
UEs, the independence among elements can be reasonably assumed. This
independence together with the sparsity in the UE domain leads to
an effective Gaussian-mixture (GM) model which well describes the
joint distributions of the channel elements. More surprisingly, empirical
results show that there are only few parameters involved in the GM
model that need to be determined. Therefore, the practical Bayes estimation
can be implemented by evaluating marginal probability density functions
via the approximate message passing (AMP) algorithm \cite{Donoho09AMP}
and learning GM parameters by means of the expectation-maximization
(EM) algorithm \cite{Vila13}. The numerical results show that this
Bayesian method can achieve a significant reduction in estimation
errors.

The assumption of channel vectors being sparse in the UE domain may
not hold when the UE dimension $KL$ is not large enough. A possible
remedy for this situation is suggested in the following. First, it
is desirable to understand if the GM model is also applicable for
modeling distributions of spare channel vectors in the angular domain.
Second, as angular-domain channels are very likely to consist of a
small number of block-wise non-zero segments resulting from few clusters
of scatterers, it is eminently reasonable to assume some dependence
among angular-domain channel elements. Hence, the distribution of
the channel vector could be a mixture of Gaussian random vectors,
and the original AMP and EM algorithms should be modified accordingly
to this new GM model.

\subsection{Discussion and Comparison }

In the previous subsections, several methods for efficient high-dimensional
CSI acquisition have been discussed for massive MIMO communications.
Table \ref{tab: Comparison} provides a brief summary of the advantages
and disadvantages of these methods. It is shown in the table that
each method utilizes a distinct sparsity structure. However, all sparsity
structures considered in massive MIMO are based on the observation
that angular-domain channels are sparse. As a result, the second-order
statistics of massive MIMO channels inherit the sparsity structure,
yielding low-rank channel covariance matrices. In addition, as sparse
channels are collectively examined, it leads to either block-sparse
or low-rank channel matrices. When the UE dimension is comparable
to the channel dimension, sparsity in the angular domain also results
in sparsity in the UE domain. On the basis of the aforementioned sparsity
structures, different sparsity-inspired methods are developed either
to reduce training overhead or to mitigate pilot contamination.

In FDD massive MIMO, without feeding back channel measurements to
the BS side, less sparsity structures are available for developing
efficient CSI acquisition methods. Despite this limitation, the weighted
$\ell_{1}$ minimization method shows that achieving further overhead
reduction is feasible if partial support information can be obtained
in advance and properly harnessed. Interestingly, by enabling the
BS to gather perfect channel measurements from its served UEs, the
joint CSI recovery method offers an effective way of utilizing sparsity
structures across multiple UEs. If the performance superiority of
this method still holds when taking rate-limited feedback channels
into account, it will establish the fact that offloading CSI acquisition
tasks to the BS is feasible and beneficial. 

With regard to TDD massive MIMO, uplink training has more sparsity
structures to utilize as high-dimensional channels are jointly recovered
at the BS side. It is worth noting that only low-rank channel covariance
matrices have been used for pilot decontamination. Other sparsity
structures such as low-rank channel matrices and sparse UE-domain
channels have not been considered for mitigating the effects of pilot
reuse. In this regard, there is still much room for innovation in
sparsity-inspired pilot decontamination. It is also worth noting that
using perfect covariance matrices of both desired channels and interference
channels in the coordinated MMSE method has drawn criticism \cite{Zhang14PilotContamination}.
It would be intriguing to assess if there exist efficient algorithms
for learning low-rank covariance matrices. If such algorithms are
developed or identified, they should be integrated into the coordinated
MMSE method.

\subsection{Implementation Issues}

Recently investigators have examined the practical implementation
of compressed sensing based algorithms for sparse channel recovery
\cite{Lofgren14,Maechler10,Maechler10GreedyAlgorithms}. Although
the design targets are channel models in the 3GPP LTE standard, several
insights that have been provided are still valuable and applicable
to realistic implementation of sparse massive MIMO channel recovery.
It has been pointed out that greedy algorithms such as OMP or matching
pursuit (MP) are more desirable from a hardware perspective. It is
because these algorithms require lower computational complexity and
lower numerical precision when compared to convex relaxation algorithms
such as basis pursuit (BP) \cite{Maechler10}. The trade-off between
hardware complexity and denoising performance of three greedy algorithms
has been characterized in \cite{Maechler10GreedyAlgorithms} and it
is indicated that the chip area overhead required to implement the
gradient pursuit (GP) algorithm can be three times larger than MP.
The power consumption is normally proportional to this area overhead.
When it comes to the design of channel recovery algorithms in FDD
massive MIMO, which are typically performed at the UE side, the issue
of hardware complexity should be carefully taken into account. On
the other hand, at the BS side, high-dimensional channels can be recovered
by more advanced algorithms such as sparse Bayesian learning or joint
CSI recovery.

\subsection{Implications of New Propagation Models}

Most existing studies have based their CSI acquisition approaches
on the conventional MIMO channel models, which may fail to capture
some unique characteristics of massive MIMO channels. For instance,
the far-field and plane wavefront assumptions no longer hold when
antenna arrays become physically larger than the Rayleigh distance
\cite{Payami12}. On the other hand, the sheer size of antenna arrays,
where different antenna elements observe varying subsets of scatterer
clusters, makes the assumption of spatial channels being wide-sense
stationary on the array axis no longer valid \cite{Gao13Model}. While
new channel models have been proposed in \cite{Wu14Model,Wu15ChanMod}
by making a more accurate spherical wavefront assumption and taking
the non-stationarities into consideration, there is still very little
understanding of how these characteristics affect the sparsity structures
of the channels in massive MIMO systems. One previous result \cite{Jiang05},
however, suggests that the spherical wavefront model does adequately
characterize the rank of the channel matrix. This implies that the
new channel models can potentially affect the SDP method which exploits
the sparsity in the form of the channel matrix rank. In addition,
the possibility that none of clusters are perceptible to some antenna
elements cannot be categorically excluded, so it indicates the possible
presence of the sparsity on the array axis. These inferences suggest
that there is abundant room for further progress in identifying utilizable
sparsity structures based on the latest models.

\section{Conclusions}

In this article, the challenges of acquiring high-dimensional CSI
in FDD/TDD massive MIMO systems have been discussed. To address these
challenges and break the curse of dimensionality, one can effectively
utilize sparsity structures that uniquely appear in massive MIMO channels.
Several state-of-the-art sparsity-inspired approaches for high-dimensional
CSI acquisition have been examined and compared in terms of the sparsity
structures being exploited, while their own advantages and disadvantages
are identified. As a result of this study, the following conclusions
can be drawn. The sparsity structures that can be harnessed are conditional
on the radio propagation environments. In TDD massive MIMO, uplink
training inherently has more sparsity structures to exploit as high-dimensional
channels are jointly recovered at the BS. On the contrary, in the
FDD mode, the desired channel is normally recovered at the UE where
utilizable sparsity structures are limited. Finally, based upon existing
approaches, we have identified the potential research problems in
need of further investigation. 

\bibliographystyle{IEEEtran}
\bibliography{FRT,MassiveMIMO,Compressed_Pilot_Decontamination}

\end{document}